\documentclass[twocolumn,english,showpacs,preprintnumbers,amsmath,amssymb,floatfix]{revtex4}
\usepackage[T1]{fontenc}
\usepackage{color}
\usepackage{array}
\usepackage{amstext}
\usepackage{graphicx}
\usepackage{esint}
\usepackage[colorlinks = true,
linkcolor = magenta,
urlcolor  = blue,
citecolor = red,
anchorcolor = blue]{hyperref}

\makeatletter

\@ifundefined{textcolor}{}
{%
	\definecolor{BLACK}{gray}{0}
	\definecolor{WHITE}{gray}{1}
	\definecolor{RED}{rgb}{1,0,0}
	\definecolor{GREEN}{rgb}{0,1,0}
	\definecolor{BLUE}{rgb}{0,0,1}
	\definecolor{CYAN}{cmyk}{1,0,0,0}
	\definecolor{MAGENTA}{cmyk}{0,1,0,0}
	\definecolor{YELLOW}{cmyk}{0,0,1,0}
}

\@ifundefined{definecolor}
{\usepackage{color}}{}
\@ifundefined{definecolor}
{\usepackage{color}}{}
\makeatother

\makeatother

\usepackage{babel}
\begin{document}
	
\title{Interference effects for
	the top quark decays $t\to b+W^+/H^+(\to\tau^+\nu_\tau)$}

\author{S. Mohammad Moosavi Nejad$^{a,b}$}
\email{mmoosavi@yazd.ac.ir}
	
\author{S. Abbaspour$^a$}

\author{R. Farashahian$^a$}

	\affiliation{$^{(a)}$Faculty of Physics, Yazd University, P.O. Box
		89195-741, Yazd, Iran\\	
       $^{(b)}$School of Particles and Accelerators,
		Institute for Research in Fundamental Sciences (IPM), P.O.Box
		19395-5531, Tehran, Iran}

	\date{\today}
	
\begin{abstract}

Applying the narrow-width approximation (NWA), we first review the NLO QCD predictions for the total decay rate of top quark considering two  unstable intermediate particles: the $W^+$-boson in the standard model (SM) of particle physics and the charged Higgs boson $H^+$ in the generic type-I and II two-Higgs-doublet models, i.e. $t\to b+W^+/H^+(\to \tau^+\nu_\tau)$. We then estimate the errors arised from this approximation at leading-order perturbation theory. Finally, we shall investigate the interference effects in the factorization  of production and decay parts of intermediate particles. We will show that  for  nearly mass-degenerate states ($m_{H^+}\approx m_{W^+}$), the correction due to the interference effect is considerable.

\end{abstract}


\maketitle

\section{Introduction}
\label{sec:intro}

Since the discovery in 1995 by the CDF and D0 experiments at the $p\bar{p}$ collider Tevatron at Fermilab, the top quark has been in or near the center of attention in high energy physics. It is still the heaviest particle of the Standard Model (SM) of elementary particle physics and its short lifetime implies that it decays before hadronization takes place. The remarkably large mass implies that the top quark couples strongly to the agents of electroweak symmetry breaking, making it both an object of interest itself and a tool to investigate that mechanism in detail.\\
The CERN Large Hadron Collider (LHC), producing a $t\bar{t}$ pair  per second, is potentially a top quark factory which allows to perform  precision tests of the SM and will enhance the sensitivity of beyond-the-SM effects in the top sector.  In this regards, a lot of theoretical works has gone into firming up the cross sections for the $t\bar{t}$ pair and the single top production at the Tevatron and the LHC, undertaken in the form of higher order QCD corrections \cite{Moch:2008ai,Moch:2008qy,Kidonakis:2008mu,Cacciari:2008zb}.  Historically, improved theoretical calculations of the top quark decay width and distributions started a long time ago. In this regard, the leading order perturbative QCD corrections to the lepton energy spectrum in the decays $t\to bW^+\to b(l^+\nu_l)$ were calculated some thirty years ago \cite{Ali:1979is}. Subsequent theoretical works leading to analytic derivations implementing the ${\cal O}(\alpha_s)$ corrections were published in \cite{Corbo:1982ah,Altarelli:1982kh} and corrected in \cite{Jezabek:1988ja}  (see also Refs.~\cite{Czarnecki:1990kv,Liu:1990py,Li:1990qf}). 
Moreover, in \cite{Fischer:2000kx,Kniehl:2012mn} the ${\cal O}(\alpha_s)$ radiative corrections to the decay rate of an unpolarized top quark is calculated where the helicities of the W-gauge boson are specified as longitudinal, transverse-plus and transverse-minus.

On the other hand, in the dominant decay mode $t\to bl^+\nu_l$ the bottom quark  hadronizes into the b-jet $X_b$ before it decays. Considering this hadronization process, in \cite{Kniehl:2012mn} the energy
distribution of bottom-flavored hadrons (B-mesons) inclusively produced in the SM decay chain  of an unpolarized top quark, i.e. $t\rightarrow bW^+\rightarrow Bl^+\nu_l+X$, is studied.  In Refs.~\cite{Nejad:2013fba,Nejad:2015pca,Nejad:2016epx,Nejad:2014sla}, the ${\cal O}(\alpha_s)$ angular distribution of energy spectrum of hadrons considering the polar and azimuthal angular correlations in the rest frame decay of a polarized top quark, i.e. $t(\uparrow)\to B+W^++X$, is studied. Furthermore, the mass effects of quarks and hadrons have been also investigated.

Charged Higgs bosons emerge in the scalar sector of several extensions of the SM and are the object of various beyond SM (BSM) searches at the  LHC. Since the SM does not include any elementary charged scalar particle, thus the experimental observation of a charged Higgs boson would necessarily be a signal for  a nontrivially extended scalar sector and  a definitive evidence of new  physics beyond the SM. In recent years, searches for charged Higgs have been  done by the ATLAS and the CMS collaborations  in proton-proton collision and numerous attempts are still in progress at the LHC.\\
Among many proposed scenarios beyond the SM  which motivate the existence of charged Higgs, a generic two-Higgs-doublet model (2HDM) \cite{Lee:1973iz,Djouadi:2005gj,Gunion} provides a greater insight of the SUSY Higgs sector without including the plethora of new particles which SUSY predicts. Within this class of
models, the Higgs sector of the SM is extended by introducing an extra doublet of complex $SU(2)_L$ Higgs scalar fields. After spontaneous symmetry breaking, the two scalar Higgs doublets $H_1$ and $H_2$ yield three physical neutral Higgs bosons (h, H, A) and a pair of charged-Higgs bosons ($H^\pm$) \cite{Djouadi:2005gj}.  Moreover, after  electroweak  symmetry  breaking  each  doublet acquires a vacuum expectation value (VEV) $\textbf{v}_i$ such that $\textbf{v}_1^2+\textbf{v}_2^2=(\sqrt{2} G_F)^{-1}$ where $G_F$ is the Fermi's constant and the $\textbf{v}_1$ and $\textbf{v}_2$ are the VEVs  of $H_1$ and $H_2$, respectively. Furthermore, it is often useful to express the parameters $\tan\beta=\textbf{v}_2/\textbf{v}_1$  as the ratio of VEVs 
and  the  neutral  sector  mixing  term  $\sin(\beta-\alpha)$. In fact, the  angles  $\alpha$ and $\beta$ govern the mixing between mass eigenstates in the CP-even sector and CP-odd/charged sectors, respectively.

The dominant production and decay modes for a charged Higgs boson depend on the value of its mass with respect to the top quark mass and can be classified into three categories \cite{Degrande:2016hyf}. Among them, light charged Higgs scenarios are defined by Higgs-boson masses smaller than the  top quark mass.
In the 2HDM, the main production mode for light charged Higgses  is through the top quark decay  $t\to bH^+$. Therefore, at the CERN LHC the light Higgses  can be searched in the subsequent decay products of the top pairs $t\bar{t}\to H^\pm H^\mp b\bar{b}$ and $t\bar{t}\to H^\pm W^\mp b\bar{b}$ when charged Higgs decays into the $\tau$-lepton and neutrino. 
In \cite{Li:1990ag,Czarnecki:1992zm}, the ${\cal O}(\alpha_s)$ QCD corrections to the hadronic decay width of a charged Higgs boson, i.e. $\Gamma(t\to bH^+)$, is calculated and in \cite{Korner:2002fx} 
the leading order contribution and the
${\cal O}(\alpha_s)$ corrections to the polarized top quark decay into $bH^+$ is computed.
In \cite{MoosaviNejad:2011yp,MoosaviNejad:2012ju}, the energy distribution of B-hadrons is investigated in the unpolarized top decays through the 2HDM scenarios, i.e. $t\to bH^+\to BH^++jets$.
In \cite{MoosaviNejad:2016aad}, in the 2HDM framework the ${\cal O}(\alpha_s)$ angular distribution of energy spectrum of B/D-mesons is studied considering the polar and the azimuthal angular correlations in the rest frame decay of a polarized top quark, i.e. $t(\uparrow)\to B/D+H^++X$ followed by $H^+\to l^+\nu_l$. 
Note that, even though current ATLAS and CMS measurements exclude a light charged Higgs for most of the parameter regions in the context of the minimal supersymmetric standard model (MSSM) scenarios, these bounds are significantly weakened in the Type II 2HDM (MSSM) once the exotic decay
channel into a lighter neutral Higgs, $H^\pm\to AW^\pm/HW^\pm$, is open. In \cite{Kling:2015uba}, the production possibility of a light charged Higgs in  top quark decay via single top or top pair production is examined with a subsequent decay as $H^\pm\to AW^\pm/HW^\pm$. There is shown that this decay mode can reach a sizable branching fraction at low $\tan\beta$ once it is kinematically permitted. 
These results show that the exotic decay channel $H^+\to AW^+/HW^+$
is  complementary to the conventional  $H^+\to \tau^+\nu_\tau$ channel considered in the current MSSM scenarios.
 
Two considerable points about all mentioned works are; firstly, in all works authors have applied the narrow width approximation (NWA) in which an intermediate gauge boson ($W^+$-boson in the SM and $H^+$-boson in the 2HDM scenario) is considered as the on-shell particle and, secondly, the contribution of interference term in total decay rate of top quarks is ignored. In this work we shall examine  how much these two approximations change the results.  Note that, an important condition limiting the applicability of the narrow width approximation, however, is the requirement that there should be no interference of the contribution of the intermediate particle for which the NWA is applied with any other close-by resonance. It should be noted that, in a general case, if the mass gap between two intermediate particles is smaller than one of their total widths, the interference term between the contributions from the two nearly mass-degenerate particles may become large. In these cases, a single resonance approach or the incoherent sum of the contributions due to two resonances  does not necessarily hold. 

This work  is organized as follows.
In Sec.~\ref{sec:one}, we calculate the Born rate of top quark decay in the SM through the direct approach and the narrow width approximation. We will also present the NLO QCD corrections to the tree-level rate of top decay.
In Sec.~\ref{two}, the same calculations will be done by working in the general 2HDM.  
In Sec.~\ref{three}, we present our results for the interference effects on the total top quark decay and show when this effect is considerable. In Sec.~\ref{four},  we summarize our conclusions.

\section{Top quark decay in the SM $t\to bW^+\to bl^+\nu_l$}
\label{sec:one}

For the Cabibbo-Kobayashi-Maskawa (CKM) quark mixing matrix \cite{Cabibbo:1963yz} one has $|V_{tb}|\approx 1$. Therefore, the top quark decays within the SM are completely dominated by the mode $t\to bW^+$ followed by $W^+\to l^+\nu_l$ where $l^+=e^+, \mu^+, \tau^+$.  Since the top quark's lifetime is much shorter than the typical strong interaction time, the top quark decay dynamics is controlled by the perturbation theory. Therefore, incorporating the QED/QCD perturbative corrections one has precise theoretical predictions for the decay width to be confronted with the experimental data. As a warm-up exercise we start to calculate the decay width of the following process
\begin{eqnarray}\label{first}
	t(p_t)\to b(p_b)+W^+(p_W)\to b(p_b)+l^+(p_l)+\nu_l(p_{\nu}),
\end{eqnarray}	
at the Born approximation. The matrix element of this process at tree level is given by
\begin{eqnarray}\label{second}
M_{Born}^{SM}\equiv M_0^{SM}&=&-\frac{g_W^2}{8}|V_{tb}|\Big(\frac{-g^{\alpha\beta}+\frac{p_W^\alpha p_W^\beta}{m_W^2}}{p_W^2-m_W^2}\Big)\nonumber\\
&&\times[\bar{u}_b(p_b)\gamma_\alpha(1-\gamma_5)u_t(p_t)]\nonumber\\
&&\times[\bar{u}_\nu(p_\nu)\gamma_\beta(1-\gamma_5)v_l(p_l)],
\end{eqnarray}
where $g_W^2=4\pi\alpha/\sin^2\theta_W=8 m_W^2 G_F/\sqrt{2}$ 
is the electroweak coupling constant, $\theta_W$ is the weak mixing angle
and $m_W$ is the $W^+$ boson mass. Since the second term in the parenthesis (\ref{second}) is proportional to the lepton mass due to the conservation of the lepton current then it can be omitted simply. Therefore, the matrix element squared reads
\begin{eqnarray}\label{threee}
|M_0^{SM}(t\to b l^+\nu_l)|^2=\big(\frac{2g_W^2|V_{tb}|}{p_W^2-m_W^2}\big)^2(p_b\cdot p_\nu)(p_l\cdot p_t),
\end{eqnarray}	
where, for the convenient scalar products in the top quark rest frame one has  $2p_b\cdot p_\nu=m_t^2+m_l^2-m_b^2-2m_t E_l$ and $p_t\cdot p_l=m_tE_l$  in which  $E_l$ is the energy of  lepton in the top quark rest frame. Technically, to obtain the matrix element squared for the polarized top decay one should replace $\sum_{s_t}u(p_t, s_t)\bar{u}(p_t, s_t)=(\displaystyle{\not}{p}_t+m_t)$ in the unpolarized Dirac string by $u(p_t, s_t)\bar{u}(p_t, s_t)=(1-\gamma_5 \displaystyle{\not}{s}_t )(\displaystyle{\not}{p}_t+m_t)/2$.

Since the main contribution to the top quark decay mode (\ref{first}) comes from the kinematic region where the $W^+$ boson is near its mass-shell,  one has to take into account its finite decay width $\Gamma_{W}$. For this reason, in Eq.~(\ref{threee}) we employ the Breit-Wigner prescription  of the $W^+$ boson propagator
for which the propagator contribution of an unstable particle of mass $M_W$ and total width $\Gamma_W$ is given by 
\begin{eqnarray}\label{born30}
\frac{1}{p_W^2-M_W^2}\to\frac{1}{p_W^2-M_W^2+iM_W\Gamma_W}.
\end{eqnarray}
Thus, the matrix element squared (\ref{threee}) reads
\begin{eqnarray}\label{threeee}
|M_0^{SM}(t\to b l^+\nu_l)|^2&=&\frac{g_W^4m_t^3|V_{tb}|^2}{(p_W^2-m_W^2)^2+m_W^2\Gamma_W^2}\nonumber\\
&&\hspace{-2cm}\times E_l\bigg\{1+(\frac{m_l}{m_t})^2-(\frac{m_b}{m_t})^2-\frac{2E_l}{m_t}\bigg\}.
\end{eqnarray}
The phase space element for the three particles final state is given by
\begin{eqnarray}\label{fourr}
dPS_3&=&(2\pi)^{-5}\frac{d^3\vec{p}_b}{2E_b}\frac{d^3\vec{p}_\nu}{2E_\nu}\frac{d^3\vec{p}_l}{2E_l}\delta^4(p_t-p_b-p_\nu-p_l)\nonumber\\
&=&\frac{1}{2^5\pi^3}dE_b dE_l. 
\end{eqnarray}
Ignoring the lepton  mass ($m_l\approx 0$), the kinematic restrictions are: $m_t/2-E_l\le E_b\le m_t/2$ and $0\le E_l\le m_t/2(1-(m_b/m_t)^2)$, where $E_b$ is the energy of b-quark. 
For the Born decay width, we use  Fermi's golden rule  
\begin{eqnarray}\label{five}
d\Gamma_0=\frac{1}{2m_t}\overline{|M_0|^2}dPS_3,
\end{eqnarray}
 where $\overline{|M_0|^2}=\sum_{spin}|M_0|^2/(1+2s_t)$ and $s_t$ stands for the top quark spin.
Thus, for the case of unpolarized top quark decay  we obtain the decay rate at the lowest-order as
\begin{eqnarray}\label{mona}
&&\Gamma_0^{SM}(t\to bl^+\nu_l)=\frac{m_t\alpha^2|V_{tb}|^2}{192\pi\sin^4\theta_W}\Bigg\{2(R-1)(1+R-2\omega)\nonumber\\
&&+\bigg[3(\omega-1)(R-\omega)+\omega\frac{\Gamma_W^2}{m_t^2}\bigg]\ln\frac{\omega \Gamma_W^2+m_t^2(R-\omega)^2}{\omega \Gamma_W^2+m_t^2(1-\omega)^2}+\nonumber\\
&&\frac{1}{m_t\sqrt{\omega}\Gamma_W}\bigg[3\omega\Gamma_W^2(1+R-2\omega)+m_t^2(1-R)^3+\nonumber\\
&&m_t^2(R-\omega)^2(R+2\omega-3)\bigg]\times\nonumber\\
&&\bigg(\tan^{-1}\frac{m_t(\omega-R)}{\sqrt{\omega}\Gamma_W}+\tan^{-1}\frac{m_t(1-\omega)}{\sqrt{\omega}\Gamma_W}\bigg)\Bigg\},
\end{eqnarray}
where we defined $R=(m_b/m_t)^2$ and $\omega=(m_W/m_t)^2$.
Concentrating on the case $l^+\nu_l=\tau^+\nu_\tau$ with $m_\tau=1.776$~GeV and taking $m_t=172.98$~GeV, $m_W=80.339$~GeV, $m_b=4.78$~GeV,  $\Gamma_W=2.085\pm 0.042$~GeV, $\sin^2\theta_W=0.2312$, $\alpha=0.0077$ and $|V_{tb}|\approx 1$ \cite{Olive:2016xmw}, one has
\begin{eqnarray}\label{bornnn}
\Gamma_0^{SM}(t\to b\tau^+\nu_\tau)=0.1543.
\end{eqnarray}
Extension of this approach to higher orders of perturbative QED/QCD is  complicated. For example, at the NLO perturbative QCD the phase space  element contains four particles including a real emitted gluon so that this leads to cumbersome computations. For this reason, in all manuscripts authors have applied  the narrow width approximation which will be described in the following.

\subsection{Narrow Width  Approximation}

The separation of a more complicated process into several subprocesses involving on-shell incoming and outgoing particles  is achieved with the help of the narrow-width approximation (NWA). This approximation  is based on the observation that the on-shell contribution  is strongly enhanced if the total width is much smaller than the mass of the particle, i.e. $\Gamma\ll M$. Here, we briefly describe how the NWA works for the decay process of top quark.

On squaring the Born matrix element (\ref{second}) and taking the Breit-Wigner prescription, one is led to the following Born contribution 
\begin{eqnarray}\label{born3}
|M_0^{SM}|^2&=&\frac{1}{(p_{W}^2-m_{W}^2)^2+(m_{W}\Gamma_{W})^2}\nonumber\\
&&\times|M^{Born}(t\to b W^+)|^2\\
&&\times|M^{Born}(W^+\to l^+\nu_l)|^2,\nonumber
\end{eqnarray}
where $|M^{Born}(t\to b W^+)|^2=2g_W^2|V_{tb}|^2(p_b\cdot p_t)$ and $|M^{Born}(W^+\to l^+\nu_l)|^2=2g_W^2(p_l\cdot p_\nu)$.
We now factorize the three body decay rate  (\ref{fourr}) into the two body rates $\Gamma(t\to bW^+)$ and $\Gamma(W^+\to l^+\nu_l)$ using the NWA for the ${W^+}$-boson for which the condition $\Gamma_{W}\ll m_{W}$ holds. 
First, we introduce the following identity
\begin{eqnarray}\label{born5}
1=\int dp_{W}^2\int\frac{d^3p_{W}}{2E_{W}}\delta^4(p_{W^+}-p_{l^+}-p_{\nu_l}),
\end{eqnarray}
therefore, the decay rate (\ref{five}) reads 
\begin{eqnarray}\label{born6}
&&d\Gamma_0=2m_{W}\int\frac{dp_{W}^2}{2\pi}\overline{|M_0^{SM}|^2}\nonumber\\
&&\times \frac{1}{2m_t}\overbrace{\Big\{\frac{d^3p_b}{(2\pi)^32E_b}\frac{d^3p_{W}}{(2\pi)^32E_{W}}(2\pi)^4\delta^4(p_t-p_b-p_{W^+})\Big\}}^{dPS_2(t\to bW^+)}\nonumber\\
&&\hspace{-0.5cm}\times \frac{1}{2m_{W}}\underbrace{\Big\{\frac{d^3p_{l}}{(2\pi)^32E_{l}}\frac{d^3p_{\nu_l}}{(2\pi)^32E_{\nu_l}}(2\pi)^4\delta^4(p_{W^+}-p_{l^+}-p_{\nu_l})\Big\}}_{dPS_2(W^+\to l^+\nu_l)}\nonumber.\\
\end{eqnarray}
Next, the phase space is nicely factorized so that by substituting  $|M_0^{SM}|^2$ (\ref{born3}), one finds
\begin{eqnarray}\label{born7}
d\Gamma_0&=&\frac{m_{W}}{\pi}\int dp_{W}^2 \frac{1}{(p_{W}^2-m_{W}^2)^2+(m_{W}\Gamma_{W})^2}\nonumber\\
&&\hspace{-1cm} \times\Big\{\frac{1}{2m_t}\Big[\overline{|M^{Born}_{t\to b W^+}|^2}\Big] dPS_2(t\to bW^+)\nonumber\\
&&\hspace{-1cm}\times\frac{1}{2m_{W}} \Big[\overline{|M^{Born}_{W^+\to l^+\nu_l}|^2}\Big] dPS_2(W^+\to l^+\nu_l)\Big\}.
\end{eqnarray}
Adopting the NWA approach, the Breit-Wigner Resonance is replaced by a delta-function as \cite{Fuchs:2014ola}
\begin{eqnarray}\label{born8}
\frac{1}{(p_{W}^2-m_{W}^2)^2+(m_{W}\Gamma_{W})^2}\approx\frac{\pi}{m_{W}\Gamma_{W}}\delta(p_{W}^2-m_{W}^2).
\end{eqnarray}
This approximation is expected to work reliably up to terms of $\mathcal{O}(\Gamma_{W}/m_{W^+})$. As was discussed, a necessary  condition limiting the applicability of this  approximation is the requirement that there should be particles with a total decay width much smaller than their mass, otherwise the integral (\ref{born7}) is reduced to \cite{Fuchs:2014ola}
\begin{eqnarray}
\int dp_{W}^2 \frac{1}{(p_{W}^2-m_{W}^2)^2+(m_{W}\Gamma_{W})^2}&=&\nonumber\\
-\frac{1}{m_{W}\Gamma_{W}}\tan^{-1}[\frac{m_{W}^2-p_{W}^2}{m_{W}\Gamma_{W}}].
\end{eqnarray}
Using the NWA, the three body decay $t\to b\tau^+\nu_\tau$  is factorized as
\begin{eqnarray}\label{born9}
\Gamma(t\to bl^+\nu_l)&=&\Gamma(t\to bW^+)\frac{\Gamma(W^+\to l^+\nu_l)}{\Gamma_{W}}\nonumber\\
&&\hspace{-0.5cm}=\Gamma(t\to bW^+) Br(W^+\to l^+\nu_l),
\end{eqnarray}
which is a result expected from physical intuition and it is expected to work reliably up to terms of ${\cal O}(\Gamma_W/m_{W^+})$.
For three individual leptonic branching ratios, one has $Br(W^+\to e^+\nu_e)=10.75\pm 0.13$, $Br(W^+\to \mu^+\nu_\mu)=10.57\pm 0.15$ and $Br(W^+\to \tau^+\nu_\tau)=11.25\pm 0.20$ in units $10^{-2}$ \cite{Beringer:1900zz}.\\
In (\ref{born9}), the partial Born width of the decay $t\to bW^+$ differential in the  angle $\theta_P$ enclosed between the top-quark
polarization three-vector $\vec{P}$ and the bottom quark three-momentum
$\vec{p}_b$, is given by  \cite{Nejad:2013fba,Nejad:2016epx}
\begin{eqnarray}
\frac{d\Gamma_0^{SM}}{d\cos\theta_P}(t\to
bW^+)=\frac{1}{2}\bigg(\Gamma_0^{SM}+P\Gamma_{0P}^{SM}\cos\theta_P\bigg),
\end{eqnarray}
where $P=|\vec{P}\,|(0\le P\le 1)$ is the degree of polarization and 
\begin{eqnarray}
&&\hspace{-0.5cm}\Gamma_0^{SM}=\frac{m_t \alpha\sqrt{S^2-R}}{8\sin^2\theta_W}(1+R-2\omega+\frac{(1-R)^2}{\omega}),\nonumber\\
&&\hspace{-0.5cm}\Gamma_{0P}^{SM}=\frac{m_t \alpha}{4\omega\sin^2\theta_W}(1-R-2\omega)(S^2-R).
\end{eqnarray}
Here, we defined $S=(1+R-\omega)/2$. Taking the input parameters as before, one has
\begin{eqnarray}
\Gamma_0^{SM}(t\to bW^+)&=&1.463,\nonumber\\
\Gamma_{0P}^{SM}(t(\uparrow)\to bW^+)&=&0.579.
\end{eqnarray}
Considering the factorization (\ref{born9}), one has $\Gamma_0^{SM}(t\to b\tau^+\nu_\tau)=1.463\times Br(W^+\to \tau^+\nu_\tau)=0.1645$ which is in agreement to the result obtained in the full calculation (\ref{bornnn}) up to the accuracy about $5\%$. 
One also has $\Gamma_0^{SM}(t\to b\mu^+\nu_\mu)=0.1543$ and $\Gamma_0^{SM}(t\to be^+\nu_e)=0.1569$.\\
In \cite{Kniehl:2012mn}, we have calculated the NLO QCD corrections to the differential decay rate of $t\to bW^+$ in the massless (with $m_b=0$) and massive (with $m_b\neq 0$) schemes. The result for the massless decay rate reads
\begin{eqnarray}
&&\Gamma_{NLO}^{SM}(t\to bW^+)=\Gamma_0^{SM}\bigg\{1+\frac{C_F\alpha_s}{2\pi}\bigg[-\frac{2\pi^2}{3}-\nonumber\\
&&4 Li_2(\omega)-\frac{5+4\omega}{1+2\omega}\ln(1-\omega)-2\frac{\omega(1+\omega)(1-2\omega)}{(1-\omega)^2(1+2\omega)}\ln\omega\nonumber\\
&&-2\ln\omega\ln(1-\omega)-\frac{1+3(1+2\omega)(\omega-2)}{2(1-\omega)(1+2\omega)}\bigg]\bigg\}.
\end{eqnarray}
In \cite{Kniehl:2012mn}, using the NWA approach we have also calculated the NLO QCD corrections to the differential decay rate of  $t\to bW^+\to bl^+\nu_l$ considering the  helicity contributions of $W^+$ boson. In \cite{Nejad:2016epx,Nejad:2014sla,Nejad:2015pca}, we have computed the  differential decay width for the process $t(\uparrow)\to bW^+$ up to the NLO accuracy. Our numerical results read
\begin{eqnarray}
\Gamma_{NLO}^{SM}(t\to bW^+)&=&\Gamma_0^{SM}(1-0.0853),\nonumber\\
\Gamma_{NLO,P}^{SM}(t(\uparrow)\to bW^+)&=&\Gamma_{0P}^{SM}(1-0.1308).
\end{eqnarray}
Thus, the ${\cal O}(\alpha_s)$ contribution to the unpolarized and polarized top width is $-8.53\%$ and $-13\%$, respectively, while the contribution of the finite W-width effect is about $5\%$. 
It should be noted that the ${\cal O}(\alpha)$ electroweak corrections contribute
typically by $+1.55\%$ \cite{Eilam:1991iz,Arbuzov:2007ke}.

\section{Top quark decay  in the 2HDM }
\label{two}

In the theories beyond the SM with an extended Higgs sector one may also have the following decay mode 
\begin{eqnarray}\label{bornn}
t\rightarrow bH^+\to bl^+\nu_l,
\end{eqnarray}
provided that $m_t>m_{H^+}+m_b$. A model independent lower bound on the Higgs mass $m_{H^+}$ arising from the nonobservation of the charged Higgs pair production at LEPII has yielded $m_{H^\pm}>79.3$~GeV at $\% 95$ C.L. \cite{Bullock:1991fd}.  As is asserted in Ref.~\cite{Ali:2009sm}, a charged Higgs with a mass in the range $80\le m_{H^\pm}\le 160$~GeV is a logical possibility and its effect should be searched for in the process (\ref{bornn}). A beginning along these lines has already been done at the Tevatron \cite{Abbott:1999eca,Abulencia:2005jd}, but a definitive search of the charged Higss bosons over a good part of the $(m_{H^\pm}-\tan\beta)$ plane is a plan  which still has to be done  and this belongs to the CERN LHC experiments \cite{Aad:2008zzm}.

Here, we review some technical detail about the decay rate of unpolarized top quark in the  process (\ref{bornn}) 
by working in the general 2HDM \cite{Lee:1973iz,Djouadi:2005gj,Gunion} where  $H_1$ and $H_2$ are the doublets  whose VEVs give masses to the down and up type quarks. 
Moreover, a linear combination of the charged components of doublets $H_1$ and $H_2$ gives two physical charged Higgs bosons $H^\pm$, i.e. $H^\pm=H_2^\pm\cos\beta-H_1^\pm\sin\beta$.\\
In a general 2HDM in order to avoid tree level flavor changing neutral currents (FCNC), that can be induced by Higgs  exchange, the generic
Higgs boson coupling to all types of quarks must be restricted. Fortunately, there  are several classes of two-Higgs-doublet models  which naturally avoid this difficulty by restricting the Higgs coupling.
Imposing flavor conservation, there are four possibilities (models I-IV) for the two Higgs doublets to couple to the SM fermions so that each gives rise to rather different phenomenology predictions.
In these four models, assuming massless neutrinos the generic charged Higgs coupling to the SM fermions
 can be expressed as a superposition of right- and left-chiral coupling factors  \cite{Barger:1989fj}, so that the relevant part of the interaction Lagrangian of the process (\ref{bornn}) is given by
\begin{eqnarray}\label{lagranj}
L_I&=&\frac{g_W}{2\sqrt{2}m_W}H^+\Big\{V_{tb}\big[\bar{u}_t(p_t)\{A(1+\gamma_5)\nonumber\\
&&\hspace{2.5cm}+B(1-\gamma_5)\}u_b(p_b)\big]
\\
&&\hspace{2.5cm}+C \big[\bar{u}_{\nu_l}(p_\nu)(1-\gamma_5)u_l(p_l)\big]\Big\},\nonumber
\end{eqnarray}
where A, B and C are three model dependent parameters.\\
In the first possibility  (called model I), the $H_2$-doublet
gives masses to all quarks and leptons so that the other one, i.e. doublet $H_1$, essentially decouples from fermions.
In this model, one has
\begin{eqnarray}\label{model1}
A_I=m_t\cot\beta , \; B_I=-m_b\cot\beta, \; C_I=-m_\tau\cot\beta.
\end{eqnarray}
In the second scenario (called model II), the  $H_2$-doublet gives mass to
the right-chiral up-type quarks (and possibly neutrinos) and  the $H_1$-doublet gives mass to the right-chiral down-type quarks and charged leptons.
In this possibility, the Lagrangian (\ref{lagranj}) contains  
\begin{eqnarray}\label{model2}
A_{II}=m_t\cot\beta, \; B_{II}=m_b\tan\beta, \; C_{II}=m_\tau\tan\beta.
\end{eqnarray}
There are also two other scenarios (models III and IV) 
in which the down-type quarks and
charged leptons receive masses from different doublets; in   model III both up- and down-type quarks couple to the second doublet ($H_2$) and all leptons to the first one, thus,  one has
\begin{eqnarray}\label{model3}
A_{III}=m_t\cot\beta, \; B_{III}=m_b\tan\beta,\;  C_{III}=-m_\tau\cot\beta\nonumber\\
\end{eqnarray}
and in the fourth scenario (model IV), the roles of  two doublets are reversed with respect to the model II, i.e. 
\begin{eqnarray}\label{model4}
A_{IV}=m_t\cot\beta, \; B_{IV}=-m_b\cot\beta,\; C_{IV}=m_\tau\tan\beta.\nonumber\\
\end{eqnarray}
These four models are also known as type I-IV 2HDM scenarios. Note that, the type-II scenario is, in fact, the Higgs sector of the MSSM up to SUSY corrections \cite{Inoue:1982pi,Fayet:1974pd}. In other words, in the MSSM we have a type-II 2HDM sector in addition to the supersymmetric particles including the  stops, charginos and gluinos.  

After this description, we start to calculate the Born term contribution to the  decay rate of the process $t\to bl^+\nu_l$ ($l^+=e^+, \mu^+, \tau^+$). Considering the decay process
\begin{eqnarray}\label{bornh}
t(p_t)\to b(p_b)+H^+(p_{H^+})\to b(p_b)+l^+(p_l)+\nu_l(p_\nu), 
\end{eqnarray}
and using the couplings from the Lagrangian (\ref{lagranj}) one can write the matrix element of the process (\ref{bornh}) as
\begin{eqnarray}\label{bornn1}
M_0^{BSM}(t\to b l^+\nu_l)&=&\frac{g_W^2 |V_{tb}|}{8m_W^2}\frac{1}{p_{H^+}^2-m_{H^+}^2+im_{H^+}\Gamma_{H}}\nonumber\\
&&\times C[\bar{u}_\nu(p_\nu)(1+\gamma_5)v_l(p_l)]\nonumber\\
&&\hspace{-1cm}\times [\bar{u}_b(p_b)\{A(1+\gamma_5)+B(1-\gamma_5)\}u_t(p_t)].\nonumber\\
\end{eqnarray}
Thus, the matrix element squared reads
\begin{eqnarray}\label{born12}
|M_0^{BSM}(t\to b l^+\nu_l)|^2&=&\nonumber\\
&&\hspace{-3.5cm}(\frac{g_W^2 |V_{tb}|}{\sqrt{2}m_W^2})^2\frac{1}{[p_{H^+}^2-m_{H^+}^2]^2+m_{H^+}^2\Gamma_{H}^2}\nonumber\\
&&\hspace{-3.5cm}\times C^2(p_l\cdot p_\nu)\{(A^2+B^2)p_b\cdot p_t+2m_b m_t AB\}.
\end{eqnarray}
The kinematic restrictions and the phase space element are as before, see Eq.~(\ref{fourr}). Thus defining  $y=(m_{H^+}/m_t)^2$, for the unpolarized decay rate one has
\begin{eqnarray}\label{mohsenn}
&&\Gamma_0^{BSM}(t\to bl^+\nu_l)=(\frac{C\alpha|V_{tb}|}{16\sqrt{\pi} m_W^2\sin^2\theta_W})^2\times\nonumber\\
&&\Bigg\{ m_t(R-1)\bigg[8AB\sqrt{R}+(A^2+B^2)(3+R-4y)\bigg]\nonumber\\
&&+2\frac{\sqrt{y}}{\Gamma_H}\bigg(\tan^{-1}\frac{m_t(y-R)}{\sqrt{y}\Gamma_H}+\tan^{-1}\frac{m_t(1-y)}{\sqrt{y}\Gamma_H}\bigg)\nonumber\\
&&\times \bigg[4AB\sqrt{R}(\Gamma_H^2+(1-y)m_t^2)+\nonumber\\
&&(A^2+B^2)\big[(2+R-3y)\Gamma_H^2+m_t^2(1-y)(1+R-y)\big]\bigg]\nonumber\\
&&+m_t\bigg[4AB(2y-1)\sqrt{R}+(A^2+B^2)\big(\frac{y\Gamma_H^2}{m_t^2}-(1+R\nonumber\\
&&-4y-2Ry+3y^2)\big)\bigg]\times\ln\frac{y\Gamma_H^2+m_t^2(R-y)^2}{y\Gamma_H^2+m_t^2(1-y)^2}\Bigg\}.
\end{eqnarray}
Leaving this result and working in the  NWA framework, where $p_{H^+}^2=m_{H^+}^2$ is put from the beginning, we have
\begin{eqnarray}\label{branchhh}
\Gamma_0^{BSM}(t\to b l^+\nu_l)=\Gamma_0(t\to bH^+)\times Br(H^+\to l^+\nu_l),\nonumber\\
\end{eqnarray}
where $Br(H^+\to l^+\nu_l)=\Gamma_0(H^+\to l^+\nu_l)/\Gamma_{H}^{\textrm{Total}}$, and the polarized and unpolarized tree-level  decay widths read \cite{MoosaviNejad:2011yp,MoosaviNejad:2012ju,MoosaviNejad:2016aad}
\begin{eqnarray}\label{unpol}
\Gamma_0^{BSM}(t\to bH^+)&=&\frac{m_t}{16\pi}\Big\{(a^2+b^2)(1+R-y)\nonumber\\
&&+2(a^2-b^2)\sqrt{R}\Big\}\lambda^{\frac{1}{2}}(1, R, y),\nonumber\\
\Gamma_{0P}^{BSM}(t(\uparrow)\to bH^+)&=&\frac{m_t}{8\pi}(ab)\lambda(1, R, y).
\end{eqnarray}
Here, $\lambda(x, y, z)=(x-y-z)^2-4yz$ is the K\"all\'en
function (triangle function) and, for simplicity, we introduced the coefficients $a$ and $b$ as
\begin{eqnarray}\label{chi}
a&=&\frac{g_W}{2\sqrt{2}m_W}|V_{tb}|(A+B),\nonumber\\
b&=&\frac{g_W}{2\sqrt{2}m_W}|V_{tb}|(A-B),\\
c&=&\frac{g_W}{2\sqrt{2}m_W}C.\nonumber
\end{eqnarray}
The advantage of this notation is that the coupling of the charged Higgs to the bottom and top quarks is expressed as a superposition of scalar and pseudoscalar coupling factors. The NLO QCD radiative corrections to the polarized and unpolarized rates  are given in our previous works  \cite{MoosaviNejad:2011yp,MoosaviNejad:2012ju,MoosaviNejad:2016aad}.\\
Since all current search strategies postulate that the charged Higgs decays either
leptonically $(H^+\to \tau^+\nu_\tau)$ or hadronically $(H^+\to c\bar{s})$, then  following Ref.~\cite{Raychaudhuri:1995kv} we adopt the relevant branching fraction $Br(H^+\to \tau^+\nu_\tau)$, as
\begin{eqnarray}\label{branchii}
Br(H^+\to \tau^+\nu_\tau)=\frac{\Gamma(H^+\to \tau^+\nu_\tau)}{\Gamma(H^+\to \tau^+\nu_\tau)+\Gamma(H^+\to c\bar{s})},
\end{eqnarray}
where, in the model I (and IV) one has
\begin{eqnarray}\label{mohs1}
&&\Gamma_0(H^+\to \tau^+\nu_\tau)=\frac{g_W^2 m_{H^+}}{32\pi m_W^2}m_\tau^2\cot^2\beta,\nonumber\\
&&\Gamma(H^+\to c\bar{s})=\frac{3g_W^2 m_{H^+}}{32\pi m_W^2}|V_{cs}|^2(\cot^2\beta)\lambda^{\frac{1}{2}}(1, \frac{m_c^2}{m_{H^+}^2}, \frac{m_s^2}{m_{H^+}^2})\nonumber\\
&&\times \Big[(m_c^2+m_s^2)(1-\frac{m_c^2}{m_{H^+}^2}-\frac{m_s^2}{m_{H^+}^2})+4\frac{m_c^2m_s^2}{m_{H^+}^2}\Big],
\end{eqnarray}
and for the model II (and III) one has
\begin{eqnarray}\label{mohs2}
&&\Gamma_0(H^+\to \tau^+\nu_\tau)=\frac{g_W^2 m_{H^+}}{32\pi m_W^2}m_\tau^2\tan^2\beta,\nonumber\\
&&\Gamma(H^+\to c\bar{s})=\frac{3g_W^2 m_{H^+}}{32\pi m_W^2}|V_{cs}|^2\lambda^{\frac{1}{2}}(1, \frac{m_c^2}{m_{H^+}^2}, \frac{m_s^2}{m_{H^+}^2})\nonumber\\
&&\times \Big[(m_c^2\cot^2\beta+m_s^2\tan^2\beta)(1-\frac{m_c^2}{m_{H^+}^2}-\frac{m_s^2}{m_{H^+}^2})\nonumber\\
&&\hspace{2cm}-4\frac{m_c^2m_s^2}{m_{H^+}^2}\Big].
\end{eqnarray}
Both results are in complete agreement with Ref.~\cite{Li:1990ag}.\\
In the limit of $m_i^2/m_H^2\to 0\; (i=c, s)$, the branching fraction  (\ref{branchii}) in the type-I 2HDM is simplified as  
\begin{eqnarray}\label{born123}
Br(H^+\to \tau^+\nu_\tau)=\frac{1}{1+3|V_{cs}|^2[(\frac{m_s}{m_\tau})^2+(\frac{m_c}{m_\tau})^2]},
\end{eqnarray}
which is independent of the $\tan\beta$ and in the type-II reads 
\begin{eqnarray}\label{born12}
Br(H^+\to \tau^+\nu_\tau)=\frac{1}{1+3|V_{cs}|^2[(\frac{m_s}{m_\tau})^2+(\frac{m_c}{m_\tau})^2\cot^4\beta]}.
\end{eqnarray}
 \begin{figure}
	\begin{center}
		\includegraphics[width=0.85\linewidth]{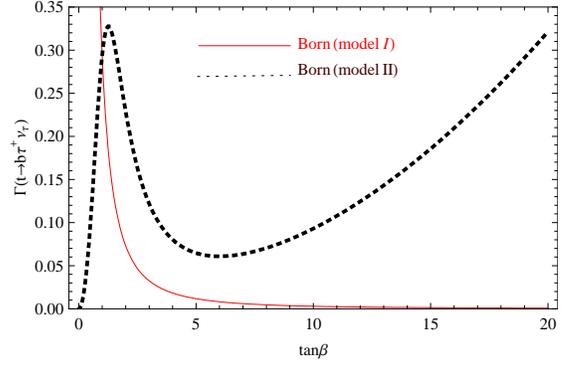}
		\caption{\label{fig1}%
			The Born decay rate of $t\to b\tau^+\nu_\tau$ as a function of $\tan\beta$ in two scenarios for which $m_{H^+}=m_{W^+}$ is set. }
	\end{center}
\end{figure}
\begin{figure}
	\begin{center}
		\includegraphics[width=0.85\linewidth]{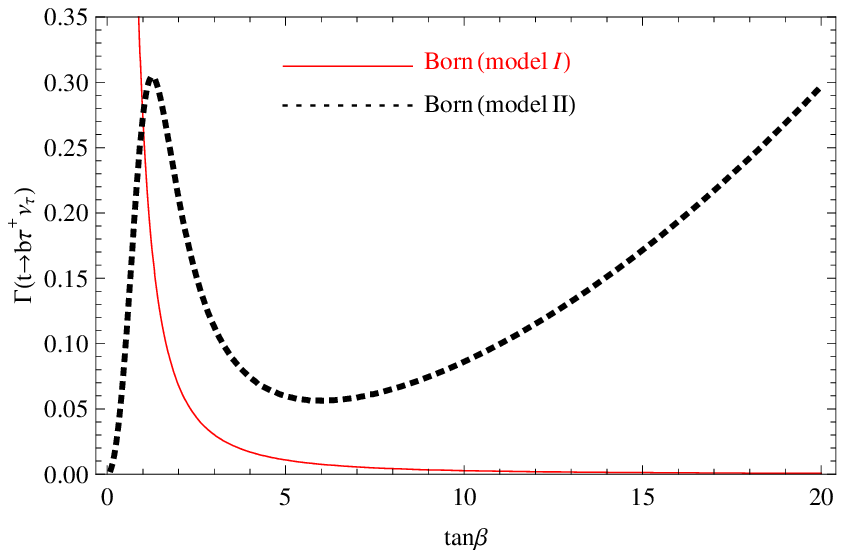}
		\caption{\label{fig12}%
			As in Fig.~\ref{fig1} but for $m_{H^+}=85$~GeV. }
	\end{center}
\end{figure}
\begin{figure}
	\begin{center}
		\includegraphics[width=0.85\linewidth]{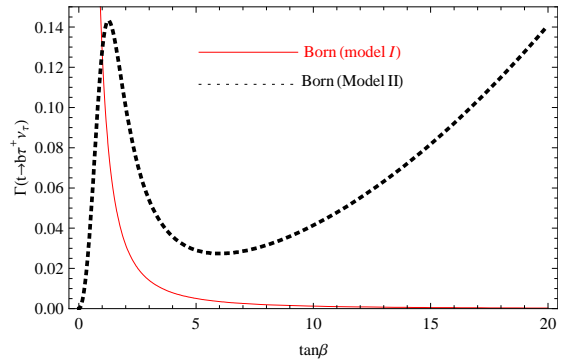}
		\caption{\label{fig2}%
			As in Fig.~\ref{fig1} but for $m_{H^+}=120$~GeV. }
	\end{center}
\end{figure}
Taking $m_s=95$~MeV, $m_c=1.67$~GeV, $m_{\tau}=1.776$~GeV and $|V_{cs}|=0.9734$, the branching ratio in the model I reads $Br=0.284$. 
While in the model I, the branching ratio  is independent of the $\tan\beta$, in the type-II scenario it depends on the $\tan\beta$. It is simple to prove that for $\tan\beta>5$  one has $Br\approx 1$ in the model II to a very high accuracy.
Direct searches at the LHC, with
the center-of-mass energy of 7 TeV \cite{Aad:2012tj,Aad:2012rjx,Aad:2013hla} and 8 TeV \cite{Khachatryan:2015uua,Khachatryan:2015qxa}
set stringent constraints on the  $m_{H^\pm}-\tan\beta$ parameter space.\\
Taking $m_{H^+}=95$~GeV and $\tan\beta=8$, from full calculation (\ref{mohsenn}) one has $\Gamma_0^{\textbf{Model I}}=36\times 10^{-4}$ in the type-I 2HDM while the corresponding result in the type-II 2HDM  reads $\Gamma_0^{\textbf{Model II}}=559\times 10^{-4}$. Considering  Eqs.~(\ref{branchhh})-(\ref{born12}), our results in the NWA scheme read: $\Gamma_0^{\textbf{Model I}}=35\times 10^{-4}$ and $\Gamma_0^{\textbf{Model II}}=568\times 10^{-4}$. As is seen, the results  (\ref{mohsenn}) obtained through the direct approach  are in good agreement with the ones from the NWA for both models up to the accuracies about $1.6\%$ (for model I) and $2.7\%$ (for model II).\\
Applying Eq.~(\ref{mohsenn}), in  Figs.~\ref{fig1}, \ref{fig12} and \ref{fig2} we studied the dependence of the top quark decay rate on $\tan\beta$ considering $m_{H^+}=m_{W^+}$ (in Fig.~\ref{fig1}), $m_{H^+}=85$~GeV (in Fig.~\ref{fig12}) and $m_{H^+}=120$~GeV (in Fig.~\ref{fig2}).  As is seen, for $\tan\beta>2$ the Born rate at the model II is always larger  than the one in the model I, and the lowest value at the type-II model occurs at $\tan\beta=6$. From Fig.~\ref{fig1}, it can be also seen that for $m_{H^+}=m_{W^+}$ one has $\Gamma_0^{\textbf{Model II}}\ge \Gamma_0^{SM}(=0.1543)$ when $\tan\beta>14$ is considered.
In  Figs.~\ref{fig3} and \ref{fig4}, varying the charged Higgs boson mass  we investigated  the behavior of top decay rate at the Born level for both models when $\tan\beta$ is fixed. 
\begin{figure}
	\begin{center}
		\includegraphics[width=0.85\linewidth]{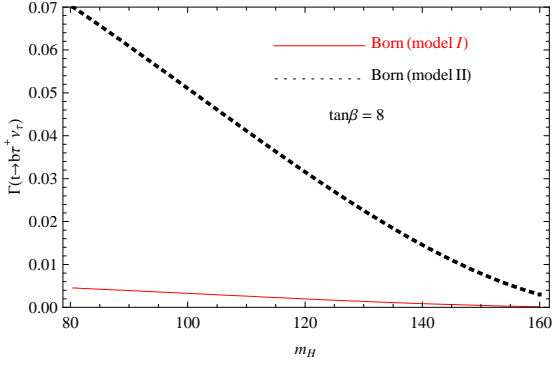}
		\caption{\label{fig3}%
			$\Gamma_{Born}^{BSM}(t\to b\tau^+\nu_\tau)$ as a function of $m_{H^+}$ in two scenarios for which $\tan\beta=8$ is set. }
	\end{center}
\end{figure}
\begin{figure}
	\begin{center}
		\includegraphics[width=0.85\linewidth]{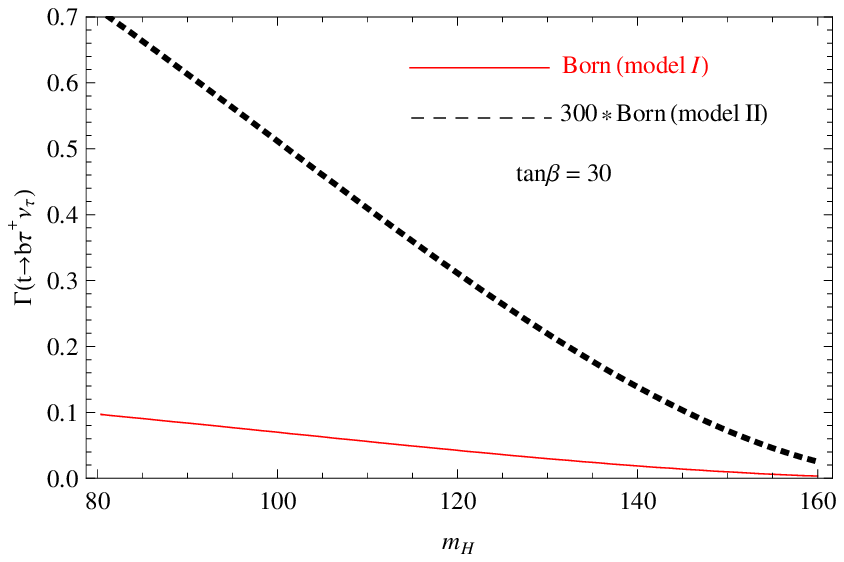}
		\caption{\label{fig4}%
			As in Fig.~\ref{fig3} but for $\tan\beta=30$. }
	\end{center}
\end{figure}
In \cite{MoosaviNejad:2012ju}, we have calculated the QCD corrections to the differential decay rate of $t\to bH^+$ up to NLO accuracy. Taking $m_{H^+}=m_{W^+}$ and $\tan\beta=8$ one has
\begin{eqnarray}
\Gamma_{NLO}^{\textbf{Model I}}(t\to bH^+)&=&\Gamma_0^{\textbf{Model I}}(1-0.0879),\nonumber\\
\Gamma_{NLO}^{\textbf{Model II}}(t\to bH^+)&=&\Gamma_0^{\textbf{Model II}}(1-0.4635),
\end{eqnarray}
where $\Gamma_0^{\textbf{Model I}}=159\times 10^{-4}$ and $\Gamma_0^{\textbf{Model II}}=700\times 10^{-4}$. Also, for $m_{H^+}=85$~GeV and $\tan\beta=8$, they read
\begin{eqnarray}
\Gamma_{NLO}^{\textbf{Model I}}(t\to bH^+)&=&\Gamma_0^{\prime\textbf{Model I}}(1-0.0870),\nonumber\\
\Gamma_{NLO}^{\textbf{Model II}}(t\to bH^+)&=&\Gamma_0^{\prime\textbf{Model II}}(1-0.4627),
\end{eqnarray}
where $\Gamma_0^{\prime\textbf{Model I}}=149\times 10^{-4}$ and $\Gamma_0^{\prime\textbf{Model II}}=656\times 10^{-4}$. As is seen the NLO QCD corrections to the decay rates are significant, specially when the type-II 2HDM scenario is concerned.

\section{Interference effects of amplitudes}
\label{three}

Considering the decay modes  
\begin{eqnarray}
t\to b+W^+/H^+(\to l^+\nu^+),
\end{eqnarray}
the full amplitude for the top decay process is the sum of the amplitudes in the SM and BSM theories, i.e. 
\begin{eqnarray}
M^{Total}(t\to b l^+\nu^+)=M_{t\to b l^+\nu^+}^{SM}+M_{t\to b l^+\nu^+}^{BSM}.
\end{eqnarray}
At the Born level, the matrix element squared is $|M_0|^2=|M_0^{SM}|^2+|M_0^{BSM}|^2+2Re(M_0^{BSM}\cdot M_0^{SM\dag})$ where  the amplitudes $M_0^{SM}$ and $M_0^{BSM}$ are given in (\ref{second}) and (\ref{bornn1}), respectively. Considering the Born  decay width (\ref{five}) and the phase space element (\ref{fourr}) one can obtain the total Born decay width as
\begin{eqnarray}\label{tot}
\Gamma_t^{Tot}=\Gamma_0^{SM}(t\to bl^+\nu_l)+\Gamma_0^{BSM}(t\to bl^+\nu_l)+\Gamma_0^{Int},
\end{eqnarray}
where $\Gamma_0^{SM}$ and $\Gamma_0^{BSM}$ are given in (\ref{mona}) and (\ref{mohsenn}), respectively. In all manuscripts it is postulated that the contribution of interference term can be ignored while it needs some subtle accuracies. In this section we intend to estimate this contribution at the leading-order, i.e. $\Gamma_0^{Int}$, to show  when one is allowed to omit this contribution.

\begin{figure}
	\begin{center}
		\includegraphics[width=0.9\linewidth]{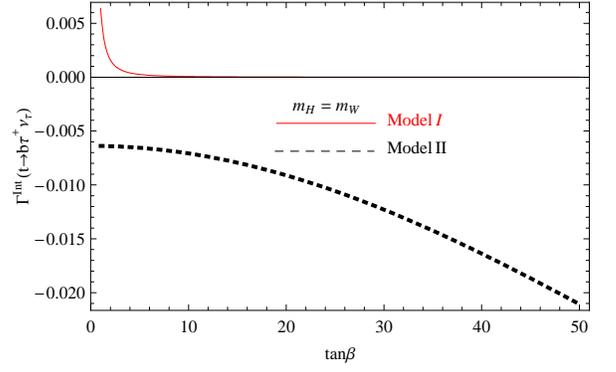}
		\caption{\label{fig5}%
			The interference contribution as a function of $\tan\beta$ for which we fixed the Higgs mass as $m_{H^+}=m_{W^+}$. }
	\end{center}
\end{figure}

\begin{figure}
	\begin{center}
		\includegraphics[width=0.9\linewidth]{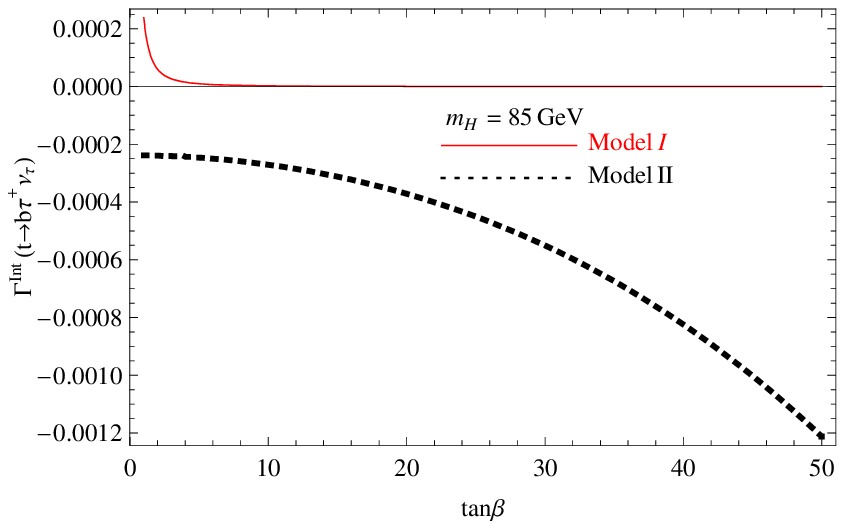}
		\caption{\label{fig67}%
			As in Fig.~\ref{fig5} but for $m_{H^+}=85$~GeV. }
	\end{center}
\end{figure}

\begin{figure}
	\begin{center}
		\includegraphics[width=0.9\linewidth]{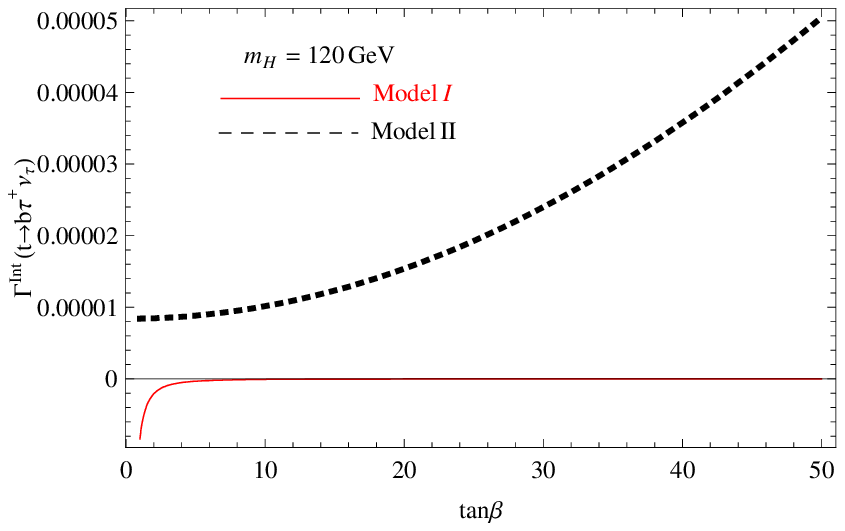}
		\caption{\label{fig6}%
			As in Fig.~\ref{fig5} but for $m_{H^+}=120$~GeV. }
	\end{center}
\end{figure}

The contribution of interference amplitude squared is obtained as 
\begin{eqnarray}
&&\overline{|M_0^{Int}|^2}=\frac{1}{2}\sum_{spin}(2Re|M_0^{BSM}.M_0^{SM\dag}|^2)=(\frac{g_W^2}{8m_W}|V_{tb}|)^2\nonumber\\
&&\times\frac{(p_W^2-m_W^2)(p_H^2-m_H^2)+m_Wm_H\Gamma_W\Gamma_H}{[(p_W^2-m_W^2)^2+m_W^2\Gamma_W^2][(p_H^2-m_H^2)^2+m_H^2\Gamma_H^2]}\times\nonumber\\
&&\big(-64m_lC[Am_t(p_b\cdot p_\nu)+Bm_b(p_t\cdot p_\nu)]\big),
\end{eqnarray}
where $p_t\cdot p_\nu=m_t E_\nu$ and $2p_b\cdot p_\nu=m_t^2+m_l^2-m_b^2-2m_tE_l$ in the top rest frame. The kinematic restrictions are
\begin{eqnarray}
&&\hspace{2.6cm} 0\le E_l\le \frac{m_t(1-R)}{2},\\
&&\frac{m_t}{2}(1-R-2\frac{E_l}{m_t})\le E_\nu\le \frac{m_t}{2}(1-\frac{m_tR}{m_t-2E_l}).\nonumber
\end{eqnarray}
Finally, defining $l=(m_l/m_t)^2$, for the contribution of interference term in the top decay rate one has
\begin{eqnarray}
&&\Gamma_0^{Int}(t\to bl^+\nu_l)=\nonumber\\
&&\frac{C\alpha^2\sqrt{l}}{2^7\pi m_t^2\omega\sin^4\theta_W}\Big\{2m_t(R-1)(A-B\sqrt{R})\nonumber\\
&&+\frac{1}{m_t^2(y-\omega)^2+(\sqrt{y}\Gamma_H+\sqrt{\omega}\Gamma_W)^2}\big[f(y,\omega,\Gamma_H,\Gamma_W)+\nonumber\\
&&g(y,\omega,\Gamma_H,\Gamma_W)+h(y,\omega,\Gamma_H,\Gamma_W)+Q(y,\omega,\Gamma_H,\Gamma_W)\big]\Big\},\nonumber\\
\end{eqnarray}
where $g(y,\omega,\Gamma_H,\Gamma_W)=f(y\leftrightarrow\omega, \Gamma_H\leftrightarrow\Gamma_W)$ so that
\begin{eqnarray}
&&f(y,\omega,\Gamma_H,\Gamma_W)=\nonumber\\
&&m_t\ln\frac{(R-y)^2m_t^2+y\Gamma_H^2}{(1-y)^2m_t^2+y\Gamma_H^2}\Bigg(m_t^2(y-\omega)\big[A(1-y)^2\nonumber\\
&&+B\sqrt{R}(1-y^2)\big]+2\sqrt{y\omega}\Gamma_H\Gamma_W\big[A(y-1)-By\sqrt{R}\big]\nonumber\\
&&+y\Gamma_H^2\big[A(y+\omega-2)-B(y+\omega)\sqrt{R}\big]\bigg),
\end{eqnarray}
and $Q(y,\omega,\Gamma_H,\Gamma_W)=h(y\leftrightarrow\omega, \Gamma_H\leftrightarrow\Gamma_W)$ where
\begin{eqnarray}
&&h(y,\omega,\Gamma_H,\Gamma_W)=\nonumber\\
&&2\bigg[\tan^{-1}\frac{m_t(R-y)}{\sqrt{y}\Gamma_H}-\tan^{-1}\frac{m_t(1-y)}{\sqrt{y}\Gamma_H}\bigg]\times\nonumber\\
&&\bigg[(B\sqrt{R}-A)\bigg(y\Gamma_W\Gamma_H^2\sqrt{\omega}+(\Gamma_H\sqrt{y})^3+\nonumber\\
&&m_t^2\Gamma_H(y^2-2y\omega+1)\sqrt{y}-m_t^2(y^2-1)\Gamma_W\sqrt{\omega}\bigg)\nonumber\\
&&+2Am_t^2\bigg((1-\omega)\Gamma_H\sqrt{y}+(1-y)\Gamma_W\sqrt{\omega}\bigg)\bigg].
\end{eqnarray}
In the above relations $\Gamma_W=2.085$ \cite{Olive:2016xmw} and concentrating on $l^+=\tau^+$ one has $\Gamma_H=\Gamma(H^+\to \tau^+\nu_\tau)+\Gamma(H^+\to c\bar{s})$ which are  given in (\ref{mohs1}) and (\ref{mohs2}) for  two models I and II.\\
In  Figs.~\ref{fig5}-\ref{fig6} we studied the dependence of the interference term on $\tan\beta$ considering $m_{H^+}=m_{W^+}$ (in Fig.~\ref{fig5}), $m_{H^+}=85$~GeV (in Fig.~\ref{fig67}) and $m_{H^+}=120$~GeV (in Fig.~\ref{fig6}).  As is seen, for $m_{H^+}=m_{W^+}$ this contribution in the model I is positive for all values of $\tan\beta$ while this is negative in the model II. This behavior is vise-versa in Fig.~\ref{fig6} where we set $m_H=120$~GeV. 
For $\tan\beta>2$, the absolute value of interference contribution at the model II is always larger  than the one in the model I.

\begin{figure}
	\begin{center}
		\includegraphics[width=0.9\linewidth]{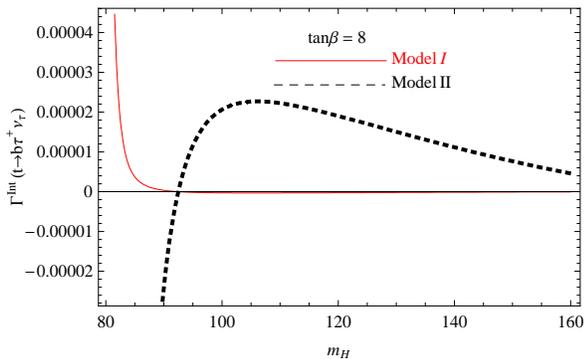}
		\caption{\label{fig7}%
			The contribution of interference term in the Born decay rate of $t\to b\tau^+\nu_\tau$ as a function of $m_{H^+}$ in two scenarios for which $\tan\beta=8$ is set.  }
	\end{center}
\end{figure}

\begin{figure}
	\begin{center}
		\includegraphics[width=0.9\linewidth]{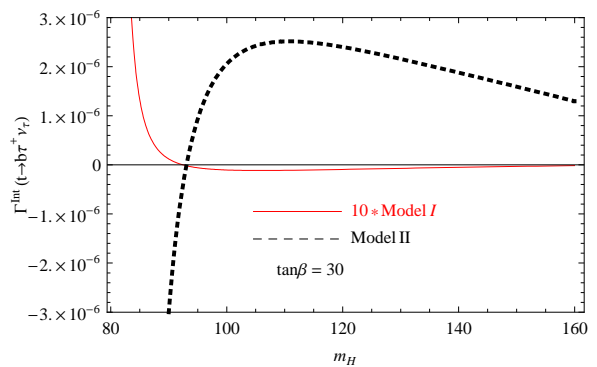}
		\caption{\label{fig8}%
			As in Fig.~\ref{fig7} but for $\tan\beta=30$. }
	\end{center}
\end{figure}

In  Figs.~\ref{fig7} and \ref{fig8} we investigated the dependence of the interference term on the charged Higgs mass by fixing $\tan\beta=8$ (in Fig.~\ref{fig7}) and $\tan\beta=30$ (in Fig.~\ref{fig8}).  As is seen the maximum value of the interference contribution occurs for $m_{H^+}\approx m_{W^+}$ and it goes to zero when $\tan\beta$ increases. 
To work out our conclusion, here, we concentrate on two following examples:
\begin{enumerate}
\item Taking $m_{H^+}\approx m_{W^+}$ and $\tan\beta=8$, in the type-I 2HDM one has 
\begin{eqnarray}
\Gamma_{NLO}^{Total}(t\to b\nu^+\nu_\tau)&=&\Gamma_0^{SM}(1-0.0853)+\nonumber\\
&&\hspace{-0.5cm}\Gamma_0^{BSM,I}(1-0.0879)+\Gamma^{Int},
\end{eqnarray}
where $\Gamma_0^{SM}=0.1543$,  $\Gamma_0^{BSM,I}=4.53\times 10^{-3}$ and the interference contribution at lowest order reads $\Gamma_0^{Int}=9.95\times 10^{-5}$. \\
This result in the type-II 2HDM reads
\begin{eqnarray}
\Gamma_{NLO}^{Total}(t\to b\nu^+\nu_\tau)&=&\Gamma_0^{SM}(1-0.0853)+\nonumber\\
&&\hspace{-0.5cm}\Gamma_0^{BSM,II}(1-0.463)+\Gamma^{Int},
\end{eqnarray}
where $\Gamma_0^{BSM,II}=6.94\times 10^{-2}$ and the interference contribution reads $\Gamma_0^{Int}=-6.82\times10^{-3}$. This example shows that the interference contribution in the type-I and II scenarios is about $2\%$ and $-9\%$ of the contribution from 2HDM at LO, respectively. 
\item Taking $m_{H^+}=85$~GeV and $\tan\beta=8$, in the type-I 2HDM one has 
\begin{eqnarray}
\Gamma_{NLO}^{Total}(t\to b\nu^+\nu_\tau)&=&\Gamma_0^{SM}(1-0.0853)+\nonumber\\
&&\hspace{-0.5cm}\Gamma_0^{BSM,I}(1-0.870)+\Gamma^{Int},
\end{eqnarray}
where $\Gamma_0^{BSM,I}=4.2\times 10^{-3}$ and the interference contribution at LO reads $\Gamma_0^{Int}=9.2\times 10^{-6}$.\\
This result in the type-II 2HDM reads
\begin{eqnarray}
\Gamma_{NLO}^{Total}(t\to b\nu^+\nu_\tau)&=&\Gamma_0^{SM}(1-0.0853)+\nonumber\\
&&\hspace{-0.6cm}\Gamma_0^{BSM,II}(1-0.462)+\Gamma^{Int},
\end{eqnarray}
where $\Gamma_0^{BSM,II}=0.0651$ and the interference term is $\Gamma_0^{Int}=-129\times 10^{-6}$.
\end{enumerate}
These two examples show that for $m_{H^+}\approx m_{W^+}$ the contribution of interference term, specifically in the type-II 2HDM,  is considerable and its value can not be ignored.  
Therefore, our numerical results emphasis that if the mass gap between two intermediate particles ($W^+$ and $H^+$ in our work) is smaller than one of their total widths, the interference term between the contributions from the two nearly mass-degenerate particles may become considerable. In other words, the interference effects can be considerable if there are several resonant diagrams whose intermediate particles (in general, with masses $M_1$ and $M_2$ for two resonances) are close in mass compared to their total decay widths: $|M_1-M_2|\leq (\Gamma_1, \Gamma_2)$ \cite{Fuchs:2014ola}. In these situations, a single resonance approach or the incoherent sum of two resonance contributions does not necessarily hold and it needs more attention. In fact, if the mass difference is smaller than their total widths, the two resonances overlap. This can lead to a considerable interference term which was neglected in the standard NWA, but can be taken into account in the full calculation or in a generalized NWA \cite{Fuchs:2014ola}.

\section{Conclusions}
\label{four}

In a general 2HDM, the main production mode of light charged Higgs bosons (with $m_{H^+}\le m_t$) is through the top quark decay, $t\to bH^+$, followed by $H^+\to \tau^+\nu_\tau$. 
In this work, we have calculated the total decay rate of top quark, i.e. $t\to b+W^+/H^+\to bl^+\nu_l$,  at the standard model of particle physics as well as the 2HDM theory. In the first part of our work,  we calculated the Born decay rate 
for the full process in which one deals with the stable intermediate bosons $W^+$ and $H^+$. Extension of this procedure to higher orders of perturbative QED/QCD is  complicated but it would be possible using   the narrow-width approximation for particles having a total width  much smaller than their masses. 
Next, using the NWA we recalculated the aforementioned decay rates and  showed that the accuracy of NWA is about $(2-5)\%$. Within this approach  we presented our numerical analysis at NLO.\\
A necessary and important condition limiting the applicability of NWA is the requirement that there should be no interference of the contribution of the intermediate particle for which the NWA is applied with any other close-by resonance. While within the SM of particle physics this condition is usually valid for relevant processes at high-energy colliders such as the CERN LHC or a future Linear Collider, many models of physics beyond the SM  have mass spectra where two or more states can be nearly mass-degenerate. If the mass gap between two intermediate particles is smaller than one of their total widths,  their resonances overlap so that the interference contribution can not be neglected if the two states mix.\\
In the last section of our paper, we investigated the interference effect of two intermediate particles in top decay, i.e. $W^+$ and $H^+$ bosons, and showed when this effect is considerable and the NWA is insufficient. Our results confirmed that when the mass of charged Higgs boson is considered  equal or near to the $W^+$-mass (referred as nearly mass-degenerate particles), the interference effects are sizable and considerable, specifically for the type-II 2HDM.
For larger values of $m_{H^+}$ this contribution can be omitted with high accuracy. \\  
It should be pointed out that several cases have been already identified in the literature in which the NWA is insufficient due to sizable interference effects, e.g. in the context of the MSSM in Refs.~\cite{Berdine:2007uv,Kalinowski:2008fk} and in the context of two- and multiple-Higgs models and in Higgsless models in Ref.~\cite{Cacciapaglia:2009ic}.

\end{document}